\documentclass[nofootinbib,prl,amsmath,amssymb,twocolumn,floatfix]{revtex4-2}

\usepackage{graphicx}
\usepackage{dcolumn}
\usepackage{bm}
\usepackage{siunitx}

\usepackage{tikz}
\usepackage{tabularx}

\newcommand{\rr}{\mathcal{R}}
\renewcommand{\tt}{\mathcal{T}}
\newcommand{\R}{\mathbb{R}}
\renewcommand{\P}{\mathcal{P}}
\newcommand{\I}{\mathrm{i}}
\renewcommand{\P}{\mathcal{P}}
\newcommand{\tr}{\mathrm{tr}}
\newcommand{\C}{\mathbb{C}}
\renewcommand{\L}{\mathcal{L}}

\begin{document}

\preprint{APS/123-QED}

\title{Graded Quasiperiodic Metamaterials Perform Fractal Rainbow Trapping}

\author{B. Davies$^{1}$, G.~J. Chaplain$^{2}$, T.~A. Starkey$^2$ and R.~V. Craster$^{1,3,4}$}
\affiliation{$^1$Department of Mathematics, Imperial College London, London SW7 2AZ, United Kingdom \\
$^2$Centre for Metamaterial Research and Innovation, Department of Physics and Astronomy, University of Exeter, Exeter EX4 4QL, United Kingdom \\
$^3$Department of Mechanical Engineering, Imperial College London, London SW7 2AZ, United Kingdom \\
$^4$UMI 2004 Abraham de Moivre-CNRS, Imperial College London, London SW7 2AZ, United Kingdom}

\date{\today}

\begin{abstract}
The rainbow trapping phenomenon of graded metamaterials can be combined with the fractal spectra of quasiperiodic waveguides to give a metamaterial that performs fractal rainbow trapping. This is achieved through a graded cut-and-project algorithm that yields a projected geometry for which the effective projection angle is graded along its length. As a result, the fractal structure of local band gaps varies with position, leading to broadband `fractal' rainbow trapping. We demonstrate this principle by designing an acoustic waveguide, which is characterised using theory, simulation and experiments.
\end{abstract}

\maketitle

\section{Introduction}

Graded metamaterials have become popular for several significant applications, such as wave energy harvesting \cite{deponti2020graded, zhao2022graded} and machine hearing \cite{rupin2019mimicking, marrocchio2021waves, ammari2023asymptotic}. They have heterogeneous micro-structures that are slowly varied (or `graded') to yield different effective properties in different spatial regions. Graded metamaterials were first developed in optics, to realise so-called ``rainbow trapping'' \cite{tsakmakidis2007trapped}. This is the phenomenon of different frequencies being reflected at different positions, due to a local effective band gap that is slowly shifted by a monotonic gradient function. This principle has since been applied in many other physical settings, such as acoustics \cite{zhu2013acoustic},  elasticity \cite{arreola2019experimental, skelton2018multi}, seismology \cite{colombi2016seismic} and water waves \cite{bennetts2019low}.

Another exciting but as-yet-unrelated direction for metamaterial physics is the move beyond periodic micro-structures into the realm of quasiperiodic metamaterials. Quasicrystals are structures that are ordered and deterministic but non-periodic. While predicting the transmission spectra of quasiperiodic systems is a notoriously challenging and longstanding problem \cite{simon1982almost, avila2009ten, kohmoto1984cantor}, they have some very appealing properties. In particular, quasiperiodic metamaterials often have a fractal structure of many large spectral gaps \cite{zolla1998remarkable, lai2002large, rechtsman2008optimized, florescu2009complete}. This work designs graded metamaterials that exploit these large gaps. 

Quasicrystals may also have some beneficial robustness properties. Their large spectral gaps are retained under periodic approximations \cite{steurer1999periodic, davies2023super}, for example. Similarly, localised modes (such as edge modes) in quasiperiodic waveguides may benefit from innate robustness with respect to imperfections \cite{marti2021edge, liu2022topological, ni2019observation}. However, it's not generally clear if this is linked to underlying topological properties \cite{simon1982almost, kraus2012topological, prodan2016bulk} or even if any robustness benefits exist when the system is appropriately normalised \cite{davies2022symmetry}.

In this Letter, we take advantage of the complex, fractal structure of many spectral gaps that is typical of a quasicrystal to produce an acoustic metamaterial that performs broadband `fractal' rainbow trapping. The quasicrystal literature contains many variants of the famous `Hofstadter butterfly' \cite{hofstadter1976energy}. These are plots that show how the spectrum varies when a parameter is modulated \cite{pal2019topological, davies2022symmetry, marti2021edge, apigo2018topological}. They typically show a self-similar and fractal collection of spectral gaps that shift up/down and open/close as the parameter is varied (sometimes resembling a butterfly). Given that such a complex collection of (sometimes large) band gaps is an appealing prospect for broadband wave control, the aim of this work is to design a graded quasiperiodic metamaterial that leverages this to perform `fractal' rainbow trapping. 

\begin{figure*}
    \centering
    \includegraphics[width = \textwidth]{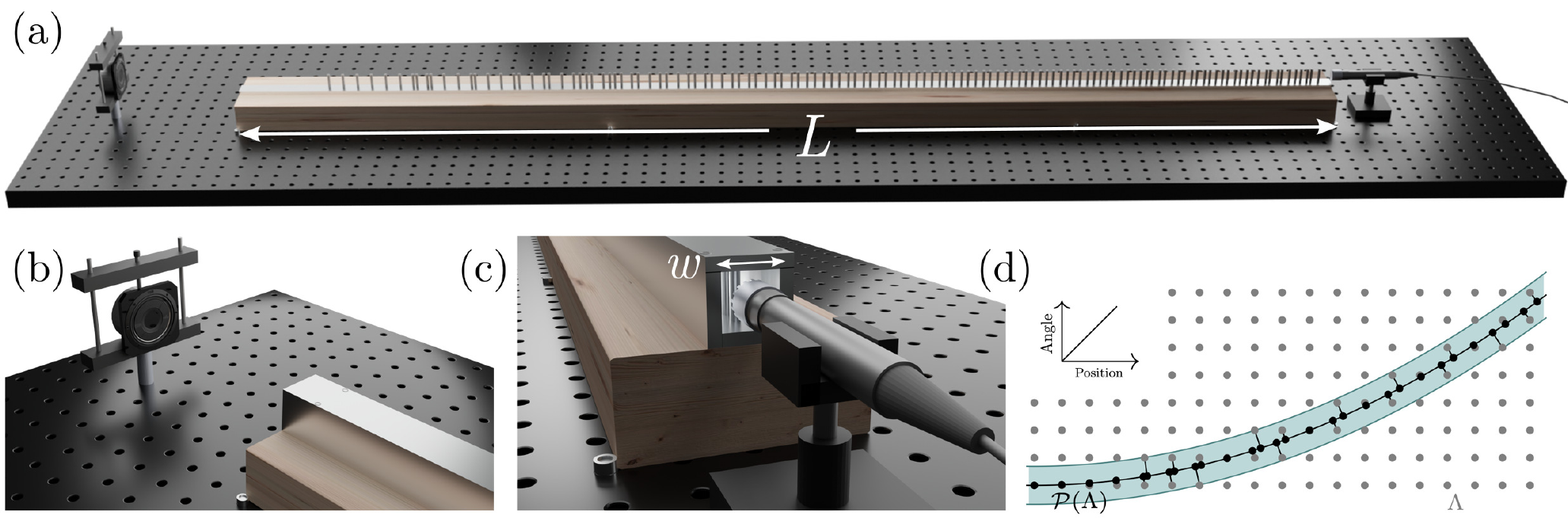}
    \caption{Experimental schematic and cut-and-project algorithm: (a) 127 acoustically rigid rods (of diameter 4 mm) are placed at positions determined by the graded cut-and-project algorithm, forming an array of length $L$ = 1.5 m, shown without the enclosing waveguide. (b,c) End-views of source (loudspeaker) and receiver (microphone) positions respectively, with enclosing waveguide shown with a square-cross section of width $w$ = 2 cm. (d) Schematic of the graded cut-and-project algorithm, which projects a square lattice $\Lambda$ onto a quadratic curve. The rod positions and sample geometry are further detailed in the supplementary material.}
    \label{fig:INeedAHero}
\end{figure*}

There have been a variety of attempts to optimise the performance of graded metamaterials \cite{wilks2022rainbow, rosafalco2022optimised, rg_exponential}. This is typically achieved by modifying the gradient function applied to a conventional periodic metamaterial, however this is a challenging problem \cite{davies2023difficulty}. Another approach has been to exploit recent breakthroughs in the development of topologically protected waveguides \cite{chaplain2020topological}. By breaking symmetries in the system, zero group velocity modes can be induced inside the Brillouin zone \cite{chaplain2020delineating}. This was shown to lead to longer interaction times and potentially increase the energy harvested from a wave. Appropriate notions of topology have been developed for quasicrystals and can be applied to the setting considered in this work \cite{simon1982almost, kraus2012topological, prodan2016bulk}. However, even if the band gaps in this work can be argued to be topologically non-trivial, the zero group velocity modes are due to Bragg conditions at the edge of the Brillouin zone. Instead, any potential performance gains resulting from to this work are thanks to the many large band gaps giving broadband effects.

The crucial feature of quasicrystals, which is central to our work, is that they can be obtained by taking incommensurate projections of higher-dimensional periodic structures \cite{janot1994quasicrystals}; the angle of this projection often serves as the canonical spectrum-modulating parameter. We will grade the quasicrystal by grading the projection angle, achieved by projecting onto a curved line (see Fig.~\ref{fig:INeedAHero}). We will use this as the basis for the design of our metamaterial, which is an acoustic waveguide with scatterers placed at the points specified by the curved cut-and-project algorithm. A wave travelling through this system experiences the fractal sequence of effective band gaps shown in Fig.~\ref{fig:butterfly} - a typical butterfly-type plot. The tendency of these gaps to shift upwards as the angle increases leads to a fractal rainbow trapping effect, which we demonstrate both theoretically and experimentally.

\section{Graded cut and project}

The quasicrystals considered in this work are one-dimensional lattices that are obtained by projecting a two-dimensional square lattice. Suppose we have a periodic lattice $\Lambda\subset\R^2$ and a curve $\Gamma\subset\R^2$. If a point in $\Lambda$ is within a distance $w>0$ of the curve $\Gamma$, we project it onto $\Gamma$ at the closest point (if this is not unique, we take all equidistant points) to give the projected lattice
\begin{equation} \label{eq:projection}
    \P(\Lambda)=\left\{ z\in\Gamma \, \bigg| \,
    \begin{matrix} 
    \exists y\in\Lambda \text{ such that } \|z-y\|_2<w \\
    \text{and }\|z-y\|_2=\min\limits_{\zeta\in\Gamma}\|\zeta-y\|_2
    \end{matrix}
    \right\}.
\end{equation}
The set $\P(\Lambda)$ is a set of points in $\Gamma\subset\R^2$ that are distributed along the curve $\Gamma$, meaning it is fundamentally a one-dimensional set embedded in $\R^2$. There are many ways to project $\P(\Lambda)$ into one dimension; for this work we choose to do so by just retaining the first coordinate of each point. This has the effect of normalising the distances with respect to the length of the curve $\Gamma$.

Predicting the transmission spectra of quasiperiodic systems is notoriously challenging and has been a longstanding problem for spectral theorists \cite{simon1982almost, avila2009ten, kohmoto1984cantor}. Taking advantage of the cut-and-project operator is a promising strategy \cite{rodriguez2008computation, amenoagbadji2023wave}, however approximating the quasicrystal by a periodic material (often known as a \emph{supercell}) remains the most prevalent approach \cite{steurer1999periodic, davies2023super}.

We use a projected quasicrystal $\P(\Lambda)$ to place scatterers within an acoustic waveguide (see Fig.~\ref{fig:INeedAHero}). If two points are close enough that the scatterers would touch or overlap, we remove the second one. A simple theoretical one-dimensional model can be used to obtain an initial theoretical characterisation of the spectrum of this system. This model is a one-dimensional scalar wave equation with a characteristic acoustic impedance that increases each time the wave passes a scatterer.

\begin{figure}
    \centering
    \includegraphics[width=\linewidth]{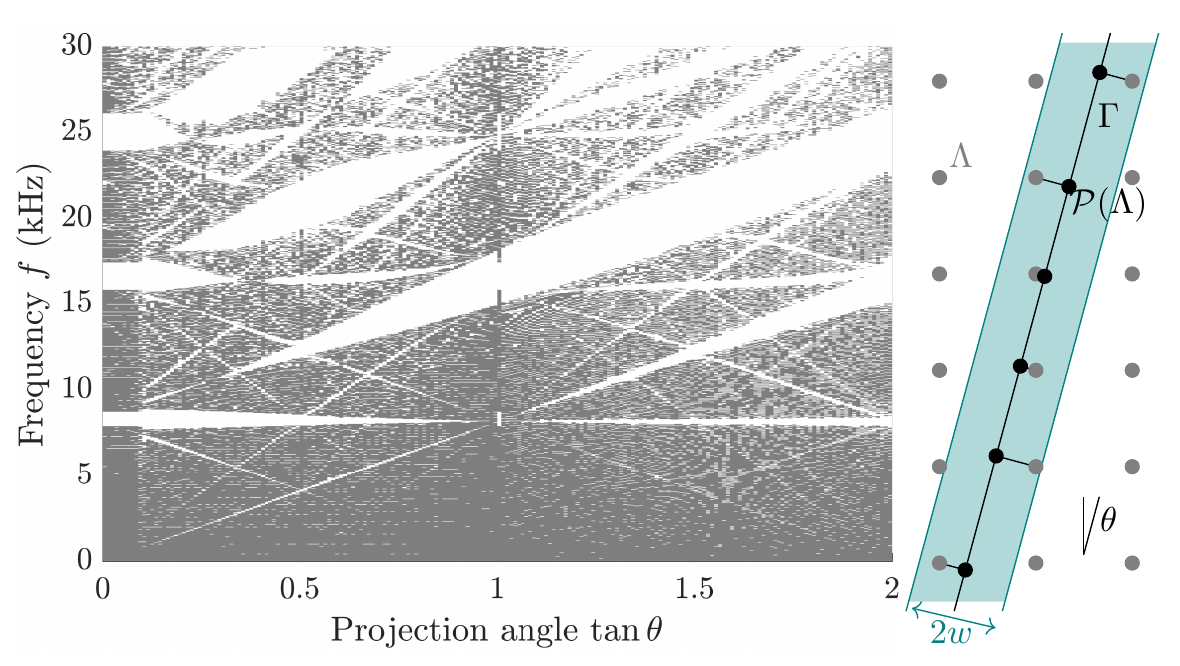}
    \caption{The spectral `butterfly'. Varying the cut-and-project angle $\theta$ of a square lattice causes the band gaps to shift and open or close. Any points in the square periodic lattice $\Lambda$ that are within a distance $w$ of the straight line $\Gamma$ are  projected onto the line, yielding the quasicrystal $\P(\Lambda)$.}
    \label{fig:butterfly}
\end{figure}

The obvious choice for the curve $\Gamma$ is a straight line, \emph{i.e.} $\Gamma:=\{y=\cos(\theta)x\}$. In Fig.~\ref{fig:butterfly} we show the spectra of the projected quasicrystal as the angle $\theta$ varies, computed using the theoretical 1D model. These are Bloch spectra, as the 200 values of $\theta$ are all chosen such that $\tan\theta=m/100$ is rational. More details on the theoretical model are given in the Supplementary Materials.

Fig.~\ref{fig:butterfly} shows a complex pattern of band gaps, that displays the self-similar properties typical of the spectral `butterflies' associated to quasiperiodic systems. The propensity for gaps to sweep upwards as $\tan\theta$ increases is reminiscent of the mechanism that leads to rainbow trapping in conventional graded metamaterials. If we design a graded quasicrystal for which the effective projection angle gradually increases along its length, then the local effective band gaps will shift upwards. This means that, at least above 12~kHz, the higher a frequency is, the further it will be able to travel before it experiences a given band gap. We achieve this grading of the projection angle by projecting onto a quadratic curve $\Gamma$, as depicted in Fig~\ref{fig:INeedAHero}(d). The choice of a quadratic means that the angle varies in direct proportion to the position in the array.

In Fig.~\ref{fig:butterfly}, gaps appear around 8~kHz, 16~kHz and 24~kHz for many different projection angles. This has been inherited from the original periodic lattice, which has band gaps at these frequencies. Since our projection algorithm \eqref{eq:projection} includes a final step that effectively normalises with respect to the length of the curve $\Gamma$, the projected quasicrystal inherits a tendency for scatterers to be distributed with the same average separation distances. Hence, we expect the graded quasicrystal to also have strong gaps at these frequencies; we detail this through additional experiments in the supplementary material.

The discontinuities that are visible in Fig.~\ref{fig:butterfly} are due to discontinuities in the number of points in $\P(\Lambda)$ that lie within a given interval. This is due to points in the square lattice $\Lambda$ leaving or entering the $2w$-wide projection strip as the angle $\theta$ varies as well as our decision to remove points that are too close to each other. This is particularly visible at $\tan\theta=1$ ($\theta=\pi/4$), where the projected crystal is very different from nearby angles (but turns out to be the same as at $\theta=0$).

\section{Experimental \& Numerical results}
We experimentally verify the fractal rainbow effect using two experimental procedures, performed using the experimental set-up shown in Fig.~\ref{fig:INeedAHero}. We form an acoustic waveguide in air, using aluminium plates to ensure sound hard boundaries, with square cross-section of width $w = 2$ cm and length $L = 1.5$ m. The waveguide consists of a bottom plate and a surrounding `hood' that encases the bottom plate (split into three sections - see supplementary material), that can be made into two configurations: `free' (no scatterers) or `scattering' (scatterers present). In the latter configuration 127 aluminium rods, of diameter $4$ mm, are placed into holes machine drilled into the bottom plate at positions dictated by the graded cut-and-project algorithm. The holes are milled to 20~\si{\micro\meter} positional precision and are of length $w+h$, where $h$ is the thickness of the bottom plate, such that there is a tight fit within the waveguide. 

\begin{figure}
    \centering
    \includegraphics[width = 0.49\textwidth]{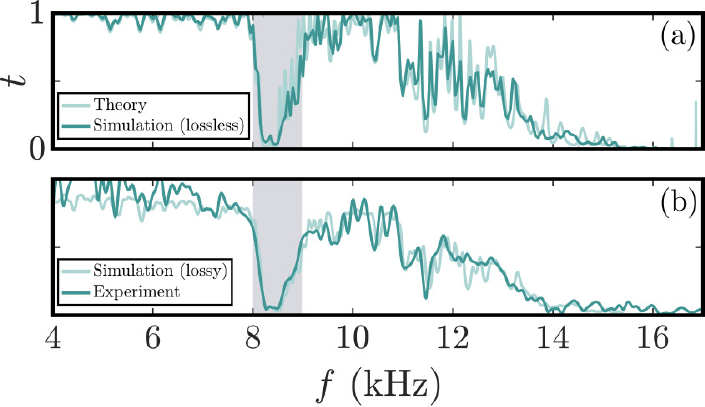}
    \caption{Comparison of the transmission spectra between (a) theory and lossless FEM simulations, and (b) Lossy FEM simulation (thermo-viscous physics) and experimental results. Note the same scales in both panels.}
    \label{fig:transmission}
\end{figure}

A transmission experiment first confirms the expected band-gap that originates at $8$~\si{\kilo\hertz}; a loudspeaker (Visaton SC 8 N) is fixed at a distance of $175$~\si{\milli\meter} from the end of the sample and emits a single-cycle Gaussian pulse, centred at 15~\si{\kilo\hertz}. A microphone (Brüel \& Kjær type 4966 $1/2$-in Free-Field, with Pre-conditioning amplifier) is placed at the exit of the waveguide, with the signal received recorded on an oscilloscope (Siglent  SDS2352X-E). Acoustic data were recorded with a sampling frequency of 5 MSa s$^{-1}$, with 50 averages taken. The experiment was conducted in both the free and scattering configurations and the spectra obtained by means of the Fast Fourier Transform (FFT). A ratio of the Fourier amplitudes with and without the scatterers then gives us a measurement of the transmission $t$. The commercial Finite Element Method (FEM) solver COMSOL Multiphysics \cite{Comsol} is used to simulate the experiment, with the problem being reduced to two dimensions. As such, the computational domain is now a rectangular waveguide ($L \times w$) with the rods being approximated by voids with sound hard boundaries. Both lossless and lossy simulations are performed, with thermoviscous physics driving the loss mechanism. Comparisons between theory and lossless simulations, and lossy simulations and experiment can be seen in Fig.~\ref{fig:transmission}. It is clear thermo-viscous losses play a role within the waveguide, and are thus adopted in all subsequent numerical models.

\begin{figure*}
    \centering
    \includegraphics[width=0.95\linewidth]{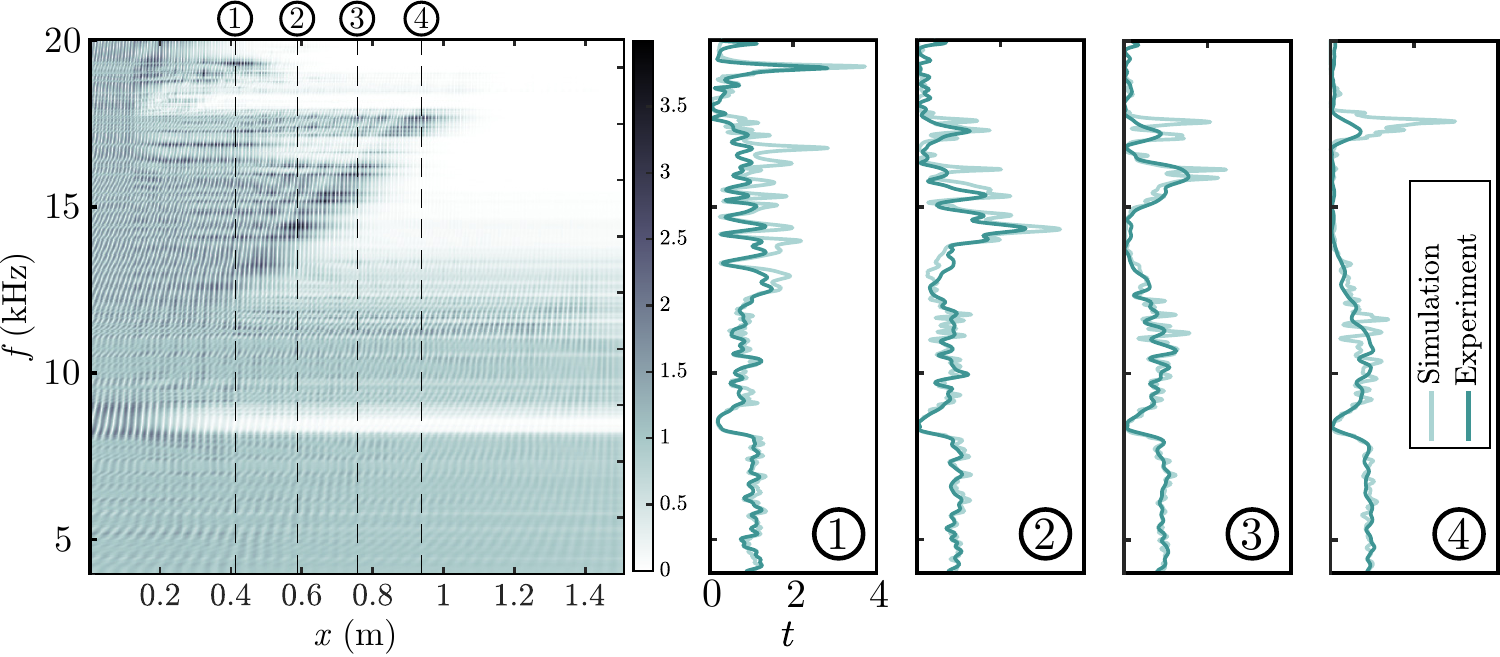}
    \caption{Demonstration of the fractal rainbow effect. FEM simulations showing frequency spectra as a function of position. The numbered lines show the measurement positions (supplementary material), with the corresponding experimental comparisons shown in the numbered plots (same frequency axis).}
    \label{fig:rainbow}
\end{figure*}
A second experiment was conducted to confirm the fractal rainbow effect: four small holes, of diameter $1.3$~\si{\milli\meter} were drilled into one side of the hood at positions $x$  = 0.41, 0.58, 0.76 0.94 m (schematic shown in the Supplementary material). A needle microphone (Brüel \& Kjær type 4182 probe microphone) was inserted into the holes, halfway between the outer walls of the waveguide and the centre lines of the rods. A repeat of the above methodology was conducted, with the ratio of the Fourier amplitude at these four positions confirming the localisation of acoustic energy as per the theoretical and numerical predictions, shown in Fig.~\ref{fig:rainbow}. Since $t$ is proportional to the ratio of pressure amplitudes, with and without the quasicrystal, its magnitude may exceed unity when the pressure field undergoes localisation by the scatterers within the waveguide. It is clear that, within the 12-18~kHz range, there is a linear relationship between the frequency and the position of reflection, demonstrating the occurrence of fractal rainbow trapping. 

\section{Conclusions}

This work shows that it is possible to combine the fields of graded metamaterials and quasicrystals to create devices that perform fractal rainbow trapping. This allows us to greatly enlarge the graded metamaterial design space and take advantage of quasicrystals' exceptional properties. In particular, the complex Cantor-set-like patterns of large spectral gaps often exhibited by quasicrystals can be exploited for graded metamaterial applications, such as energy harvesting. 

Our results pose several exciting open questions. In this work, we projected a square lattice onto a quadratic curve, giving a linear relationship between effective projection angle and position; can the performance be optimised by tuning this relationship? Given the subtleties of drawing calibrated comparisons between graded metamaterials, \cite{davies2023difficulty}, this requires a careful systematic study. As does the question of comparing the performance of graded quasiperiodic metamaterials with their conventional periodic counterparts. Finally, it remains to be seen if the topologically protected edge modes that have been observed in Harper-type quasicrystals \cite{ni2019observation,pal2019topological,apigo2018topological} can be exploited to perform topological rainbow trapping \cite{chaplain2020topological} in quasicrystals. Just by leveraging the quasicrystals whose spectra are already well understood, we can unlock many new graded metamaterials very quickly.

\begin{acknowledgments}
B.D. is supported by the Engineering and Physical Sciences Research Council through a Research Fellowship with grant number EP/X027422/1. G.J.C. gratefully acknowledges financial support from the Royal Commission for the Exhibition of 1851 in the form of a Research Fellowship. T.A.S. gratefully acknowledges financial support from Dstl.
\end{acknowledgments}

\bibliography{references}

\begin{thebibliography}{39}%
\makeatletter
\providecommand \@ifxundefined [1]{%
 \@ifx{#1\undefined}
}%
\providecommand \@ifnum [1]{%
 \ifnum #1\expandafter \@firstoftwo
 \else \expandafter \@secondoftwo
 \fi
}%
\providecommand \@ifx [1]{%
 \ifx #1\expandafter \@firstoftwo
 \else \expandafter \@secondoftwo
 \fi
}%
\providecommand \natexlab [1]{#1}%
\providecommand \enquote  [1]{``#1''}%
\providecommand \bibnamefont  [1]{#1}%
\providecommand \bibfnamefont [1]{#1}%
\providecommand \citenamefont [1]{#1}%
\providecommand \href@noop [0]{\@secondoftwo}%
\providecommand \href [0]{\begingroup \@sanitize@url \@href}%
\providecommand \@href[1]{\@@startlink{#1}\@@href}%
\providecommand \@@href[1]{\endgroup#1\@@endlink}%
\providecommand \@sanitize@url [0]{\catcode `\\12\catcode `\$12\catcode
  `\&12\catcode `\#12\catcode `\^12\catcode `\_12\catcode `\%12\relax}%
\providecommand \@@startlink[1]{}%
\providecommand \@@endlink[0]{}%
\providecommand \url  [0]{\begingroup\@sanitize@url \@url }%
\providecommand \@url [1]{\endgroup\@href {#1}{\urlprefix }}%
\providecommand \urlprefix  [0]{URL }%
\providecommand \Eprint [0]{\href }%
\providecommand \doibase [0]{https://doi.org/}%
\providecommand \selectlanguage [0]{\@gobble}%
\providecommand \bibinfo  [0]{\@secondoftwo}%
\providecommand \bibfield  [0]{\@secondoftwo}%
\providecommand \translation [1]{[#1]}%
\providecommand \BibitemOpen [0]{}%
\providecommand \bibitemStop [0]{}%
\providecommand \bibitemNoStop [0]{.\EOS\space}%
\providecommand \EOS [0]{\spacefactor3000\relax}%
\providecommand \BibitemShut  [1]{\csname bibitem#1\endcsname}%
\let\auto@bib@innerbib\@empty
\bibitem [{\citenamefont {De~Ponti}\ \emph {et~al.}(2020)\citenamefont
  {De~Ponti}, \citenamefont {Colombi}, \citenamefont {Ardito}, \citenamefont
  {Braghin}, \citenamefont {Corigliano},\ and\ \citenamefont
  {Craster}}]{deponti2020graded}%
  \BibitemOpen
  \bibfield  {author} {\bibinfo {author} {\bibfnamefont {J.~M.}\ \bibnamefont
  {De~Ponti}}, \bibinfo {author} {\bibfnamefont {A.}~\bibnamefont {Colombi}},
  \bibinfo {author} {\bibfnamefont {R.}~\bibnamefont {Ardito}}, \bibinfo
  {author} {\bibfnamefont {F.}~\bibnamefont {Braghin}}, \bibinfo {author}
  {\bibfnamefont {A.}~\bibnamefont {Corigliano}},\ and\ \bibinfo {author}
  {\bibfnamefont {R.~V.}\ \bibnamefont {Craster}},\ }\bibfield  {title}
  {\bibinfo {title} {Graded elastic metasurface for enhanced energy
  harvesting},\ }\href@noop {} {\bibfield  {journal} {\bibinfo  {journal} {New
  J. Phys.}\ }\textbf {\bibinfo {volume} {22}},\ \bibinfo {pages} {013013}
  (\bibinfo {year} {2020})}\BibitemShut {NoStop}%
\bibitem [{\citenamefont {Zhao}\ \emph {et~al.}(2022)\citenamefont {Zhao},
  \citenamefont {Thomsen}, \citenamefont {De~Ponti}, \citenamefont {Riva},
  \citenamefont {Van~Damme}, \citenamefont {Bergamini}, \citenamefont
  {Chatzi},\ and\ \citenamefont {Colombi}}]{zhao2022graded}%
  \BibitemOpen
  \bibfield  {author} {\bibinfo {author} {\bibfnamefont {B.}~\bibnamefont
  {Zhao}}, \bibinfo {author} {\bibfnamefont {H.~R.}\ \bibnamefont {Thomsen}},
  \bibinfo {author} {\bibfnamefont {J.~M.}\ \bibnamefont {De~Ponti}}, \bibinfo
  {author} {\bibfnamefont {E.}~\bibnamefont {Riva}}, \bibinfo {author}
  {\bibfnamefont {B.}~\bibnamefont {Van~Damme}}, \bibinfo {author}
  {\bibfnamefont {A.}~\bibnamefont {Bergamini}}, \bibinfo {author}
  {\bibfnamefont {E.}~\bibnamefont {Chatzi}},\ and\ \bibinfo {author}
  {\bibfnamefont {A.}~\bibnamefont {Colombi}},\ }\bibfield  {title} {\bibinfo
  {title} {A graded metamaterial for broadband and high-capability
  piezoelectric energy harvesting},\ }\href@noop {} {\bibfield  {journal}
  {\bibinfo  {journal} {Energy Convers. Manage.}\ }\textbf {\bibinfo {volume}
  {269}},\ \bibinfo {pages} {116056} (\bibinfo {year} {2022})}\BibitemShut
  {NoStop}%
\bibitem [{\citenamefont {Rupin}\ \emph {et~al.}(2019)\citenamefont {Rupin},
  \citenamefont {Lerosey}, \citenamefont {de~Rosny},\ and\ \citenamefont
  {Lemoult}}]{rupin2019mimicking}%
  \BibitemOpen
  \bibfield  {author} {\bibinfo {author} {\bibfnamefont {M.}~\bibnamefont
  {Rupin}}, \bibinfo {author} {\bibfnamefont {G.}~\bibnamefont {Lerosey}},
  \bibinfo {author} {\bibfnamefont {J.}~\bibnamefont {de~Rosny}},\ and\
  \bibinfo {author} {\bibfnamefont {F.}~\bibnamefont {Lemoult}},\ }\bibfield
  {title} {\bibinfo {title} {Mimicking the cochlea with an active acoustic
  metamaterial},\ }\href@noop {} {\bibfield  {journal} {\bibinfo  {journal}
  {New J. Phys.}\ }\textbf {\bibinfo {volume} {21}},\ \bibinfo {pages} {093012}
  (\bibinfo {year} {2019})}\BibitemShut {NoStop}%
\bibitem [{\citenamefont {Marrocchio}\ \emph {et~al.}(2021)\citenamefont
  {Marrocchio}, \citenamefont {Karlos},\ and\ \citenamefont
  {Elliott}}]{marrocchio2021waves}%
  \BibitemOpen
  \bibfield  {author} {\bibinfo {author} {\bibfnamefont {R.}~\bibnamefont
  {Marrocchio}}, \bibinfo {author} {\bibfnamefont {A.}~\bibnamefont {Karlos}},\
  and\ \bibinfo {author} {\bibfnamefont {S.}~\bibnamefont {Elliott}},\
  }\bibfield  {title} {\bibinfo {title} {Waves in the cochlea and in acoustic
  rainbow sensors},\ }\href@noop {} {\bibfield  {journal} {\bibinfo  {journal}
  {Wave Motion}\ }\textbf {\bibinfo {volume} {106}},\ \bibinfo {pages} {102808}
  (\bibinfo {year} {2021})}\BibitemShut {NoStop}%
\bibitem [{\citenamefont {Ammari}\ and\ \citenamefont
  {Davies}(2023)}]{ammari2023asymptotic}%
  \BibitemOpen
  \bibfield  {author} {\bibinfo {author} {\bibfnamefont {H.}~\bibnamefont
  {Ammari}}\ and\ \bibinfo {author} {\bibfnamefont {B.}~\bibnamefont
  {Davies}},\ }\bibfield  {title} {\bibinfo {title} {Asymptotic links between
  signal processing, acoustic metamaterials, and biology},\ }\href@noop {}
  {\bibfield  {journal} {\bibinfo  {journal} {SIAM J. Imaging Sci.}\ }\textbf
  {\bibinfo {volume} {16}},\ \bibinfo {pages} {64} (\bibinfo {year}
  {2023})}\BibitemShut {NoStop}%
\bibitem [{\citenamefont {Tsakmakidis}\ \emph {et~al.}(2007)\citenamefont
  {Tsakmakidis}, \citenamefont {Boardman},\ and\ \citenamefont
  {Hess}}]{tsakmakidis2007trapped}%
  \BibitemOpen
  \bibfield  {author} {\bibinfo {author} {\bibfnamefont {K.~L.}\ \bibnamefont
  {Tsakmakidis}}, \bibinfo {author} {\bibfnamefont {A.~D.}\ \bibnamefont
  {Boardman}},\ and\ \bibinfo {author} {\bibfnamefont {O.}~\bibnamefont
  {Hess}},\ }\bibfield  {title} {\bibinfo {title} {‘trapped rainbow’
  storage of light in metamaterials},\ }\href@noop {} {\bibfield  {journal}
  {\bibinfo  {journal} {Nature}\ }\textbf {\bibinfo {volume} {450}},\ \bibinfo
  {pages} {397} (\bibinfo {year} {2007})}\BibitemShut {NoStop}%
\bibitem [{\citenamefont {Zhu}\ \emph {et~al.}(2013)\citenamefont {Zhu},
  \citenamefont {Chen}, \citenamefont {Zhu}, \citenamefont {Garcia-Vidal},
  \citenamefont {Yin}, \citenamefont {Zhang},\ and\ \citenamefont
  {Zhang}}]{zhu2013acoustic}%
  \BibitemOpen
  \bibfield  {author} {\bibinfo {author} {\bibfnamefont {J.}~\bibnamefont
  {Zhu}}, \bibinfo {author} {\bibfnamefont {Y.}~\bibnamefont {Chen}}, \bibinfo
  {author} {\bibfnamefont {X.}~\bibnamefont {Zhu}}, \bibinfo {author}
  {\bibfnamefont {F.~J.}\ \bibnamefont {Garcia-Vidal}}, \bibinfo {author}
  {\bibfnamefont {X.}~\bibnamefont {Yin}}, \bibinfo {author} {\bibfnamefont
  {W.}~\bibnamefont {Zhang}},\ and\ \bibinfo {author} {\bibfnamefont
  {X.}~\bibnamefont {Zhang}},\ }\bibfield  {title} {\bibinfo {title} {Acoustic
  rainbow trapping},\ }\href@noop {} {\bibfield  {journal} {\bibinfo  {journal}
  {Sci. Rep.}\ }\textbf {\bibinfo {volume} {3}},\ \bibinfo {pages} {1}
  (\bibinfo {year} {2013})}\BibitemShut {NoStop}%
\bibitem [{\citenamefont {Arreola-Lucas}\ \emph {et~al.}(2019)\citenamefont
  {Arreola-Lucas}, \citenamefont {Baez}, \citenamefont {Cervera}, \citenamefont
  {Climente}, \citenamefont {M{\'e}ndez-S{\'a}nchez},\ and\ \citenamefont
  {S{\'a}nchez-Dehesa}}]{arreola2019experimental}%
  \BibitemOpen
  \bibfield  {author} {\bibinfo {author} {\bibfnamefont {A.}~\bibnamefont
  {Arreola-Lucas}}, \bibinfo {author} {\bibfnamefont {G.}~\bibnamefont {Baez}},
  \bibinfo {author} {\bibfnamefont {F.}~\bibnamefont {Cervera}}, \bibinfo
  {author} {\bibfnamefont {A.}~\bibnamefont {Climente}}, \bibinfo {author}
  {\bibfnamefont {R.~A.}\ \bibnamefont {M{\'e}ndez-S{\'a}nchez}},\ and\
  \bibinfo {author} {\bibfnamefont {J.}~\bibnamefont {S{\'a}nchez-Dehesa}},\
  }\bibfield  {title} {\bibinfo {title} {Experimental evidence of rainbow
  trapping and {Bloch} oscillations of torsional waves in chirped metallic
  beams},\ }\href@noop {} {\bibfield  {journal} {\bibinfo  {journal} {Sci.
  Rep.}\ }\textbf {\bibinfo {volume} {9}},\ \bibinfo {pages} {1} (\bibinfo
  {year} {2019})}\BibitemShut {NoStop}%
\bibitem [{\citenamefont {Skelton}\ \emph {et~al.}(2018)\citenamefont
  {Skelton}, \citenamefont {Craster}, \citenamefont {Colombi},\ and\
  \citenamefont {Colquitt}}]{skelton2018multi}%
  \BibitemOpen
  \bibfield  {author} {\bibinfo {author} {\bibfnamefont {E.~A.}\ \bibnamefont
  {Skelton}}, \bibinfo {author} {\bibfnamefont {R.~V.}\ \bibnamefont
  {Craster}}, \bibinfo {author} {\bibfnamefont {A.}~\bibnamefont {Colombi}},\
  and\ \bibinfo {author} {\bibfnamefont {D.~J.}\ \bibnamefont {Colquitt}},\
  }\bibfield  {title} {\bibinfo {title} {The multi-physics metawedge: graded
  arrays on fluid-loaded elastic plates and the mechanical analogues of rainbow
  trapping and mode conversion},\ }\href@noop {} {\bibfield  {journal}
  {\bibinfo  {journal} {New J. Phys.}\ }\textbf {\bibinfo {volume} {20}},\
  \bibinfo {pages} {053017} (\bibinfo {year} {2018})}\BibitemShut {NoStop}%
\bibitem [{\citenamefont {Colombi}\ \emph {et~al.}(2016)\citenamefont
  {Colombi}, \citenamefont {Colquitt}, \citenamefont {Roux}, \citenamefont
  {Guenneau},\ and\ \citenamefont {Craster}}]{colombi2016seismic}%
  \BibitemOpen
  \bibfield  {author} {\bibinfo {author} {\bibfnamefont {A.}~\bibnamefont
  {Colombi}}, \bibinfo {author} {\bibfnamefont {D.}~\bibnamefont {Colquitt}},
  \bibinfo {author} {\bibfnamefont {P.}~\bibnamefont {Roux}}, \bibinfo {author}
  {\bibfnamefont {S.}~\bibnamefont {Guenneau}},\ and\ \bibinfo {author}
  {\bibfnamefont {R.~V.}\ \bibnamefont {Craster}},\ }\bibfield  {title}
  {\bibinfo {title} {A seismic metamaterial: The resonant metawedge},\
  }\href@noop {} {\bibfield  {journal} {\bibinfo  {journal} {Sci. Rep.}\
  }\textbf {\bibinfo {volume} {6}},\ \bibinfo {pages} {27717} (\bibinfo {year}
  {2016})}\BibitemShut {NoStop}%
\bibitem [{\citenamefont {Bennetts}\ \emph {et~al.}(2019)\citenamefont
  {Bennetts}, \citenamefont {Peter},\ and\ \citenamefont
  {Craster}}]{bennetts2019low}%
  \BibitemOpen
  \bibfield  {author} {\bibinfo {author} {\bibfnamefont {L.~G.}\ \bibnamefont
  {Bennetts}}, \bibinfo {author} {\bibfnamefont {M.~A.}\ \bibnamefont
  {Peter}},\ and\ \bibinfo {author} {\bibfnamefont {R.~V.}\ \bibnamefont
  {Craster}},\ }\bibfield  {title} {\bibinfo {title} {Low-frequency wave-energy
  amplification in graded two-dimensional resonator arrays},\ }\href@noop {}
  {\bibfield  {journal} {\bibinfo  {journal} {Phil. Trans. R. Soc. A}\ }\textbf
  {\bibinfo {volume} {377}},\ \bibinfo {pages} {20190104} (\bibinfo {year}
  {2019})}\BibitemShut {NoStop}%
\bibitem [{\citenamefont {Simon}(1982)}]{simon1982almost}%
  \BibitemOpen
  \bibfield  {author} {\bibinfo {author} {\bibfnamefont {B.}~\bibnamefont
  {Simon}},\ }\bibfield  {title} {\bibinfo {title} {Almost periodic
  {Schr{\"o}dinger} operators: a review},\ }\href@noop {} {\bibfield  {journal}
  {\bibinfo  {journal} {Adv. Appl. Math.}\ }\textbf {\bibinfo {volume} {3}},\
  \bibinfo {pages} {463} (\bibinfo {year} {1982})}\BibitemShut {NoStop}%
\bibitem [{\citenamefont {Avila}\ and\ \citenamefont
  {Jitomirskaya}(2009)}]{avila2009ten}%
  \BibitemOpen
  \bibfield  {author} {\bibinfo {author} {\bibfnamefont {A.}~\bibnamefont
  {Avila}}\ and\ \bibinfo {author} {\bibfnamefont {S.}~\bibnamefont
  {Jitomirskaya}},\ }\bibfield  {title} {\bibinfo {title} {The ten martini
  problem},\ }\href@noop {} {\bibfield  {journal} {\bibinfo  {journal} {Ann.
  Math.}\ }\textbf {\bibinfo {volume} {170}},\ \bibinfo {pages} {303} (\bibinfo
  {year} {2009})}\BibitemShut {NoStop}%
\bibitem [{\citenamefont {Kohmoto}\ and\ \citenamefont
  {Oono}(1984)}]{kohmoto1984cantor}%
  \BibitemOpen
  \bibfield  {author} {\bibinfo {author} {\bibfnamefont {M.}~\bibnamefont
  {Kohmoto}}\ and\ \bibinfo {author} {\bibfnamefont {Y.}~\bibnamefont {Oono}},\
  }\bibfield  {title} {\bibinfo {title} {Cantor spectrum for an almost periodic
  {Schr{\"o}dinger} equation and a dynamical map},\ }\href@noop {} {\bibfield
  {journal} {\bibinfo  {journal} {Phys. Lett. A}\ }\textbf {\bibinfo {volume}
  {102}},\ \bibinfo {pages} {145} (\bibinfo {year} {1984})}\BibitemShut
  {NoStop}%
\bibitem [{\citenamefont {Zolla}\ \emph {et~al.}(1998)\citenamefont {Zolla},
  \citenamefont {Felbacq},\ and\ \citenamefont {Guizal}}]{zolla1998remarkable}%
  \BibitemOpen
  \bibfield  {author} {\bibinfo {author} {\bibfnamefont {F.}~\bibnamefont
  {Zolla}}, \bibinfo {author} {\bibfnamefont {D.}~\bibnamefont {Felbacq}},\
  and\ \bibinfo {author} {\bibfnamefont {B.}~\bibnamefont {Guizal}},\
  }\bibfield  {title} {\bibinfo {title} {A remarkable diffractive property of
  photonic quasi-crystals},\ }\href@noop {} {\bibfield  {journal} {\bibinfo
  {journal} {Opt. Commun.}\ }\textbf {\bibinfo {volume} {148}},\ \bibinfo
  {pages} {6} (\bibinfo {year} {1998})}\BibitemShut {NoStop}%
\bibitem [{\citenamefont {Lai}\ \emph {et~al.}(2002)\citenamefont {Lai},
  \citenamefont {Zhang},\ and\ \citenamefont {Zhang}}]{lai2002large}%
  \BibitemOpen
  \bibfield  {author} {\bibinfo {author} {\bibfnamefont {Y.}~\bibnamefont
  {Lai}}, \bibinfo {author} {\bibfnamefont {X.}~\bibnamefont {Zhang}},\ and\
  \bibinfo {author} {\bibfnamefont {Z.-Q.}\ \bibnamefont {Zhang}},\ }\bibfield
  {title} {\bibinfo {title} {Large sonic band gaps in 12-fold quasicrystals},\
  }\href@noop {} {\bibfield  {journal} {\bibinfo  {journal} {J. Appl. Phys.}\
  }\textbf {\bibinfo {volume} {91}},\ \bibinfo {pages} {6191} (\bibinfo {year}
  {2002})}\BibitemShut {NoStop}%
\bibitem [{\citenamefont {Rechtsman}\ \emph {et~al.}(2008)\citenamefont
  {Rechtsman}, \citenamefont {Jeong}, \citenamefont {Chaikin}, \citenamefont
  {Torquato},\ and\ \citenamefont {Steinhardt}}]{rechtsman2008optimized}%
  \BibitemOpen
  \bibfield  {author} {\bibinfo {author} {\bibfnamefont {M.~C.}\ \bibnamefont
  {Rechtsman}}, \bibinfo {author} {\bibfnamefont {H.-C.}\ \bibnamefont
  {Jeong}}, \bibinfo {author} {\bibfnamefont {P.~M.}\ \bibnamefont {Chaikin}},
  \bibinfo {author} {\bibfnamefont {S.}~\bibnamefont {Torquato}},\ and\
  \bibinfo {author} {\bibfnamefont {P.~J.}\ \bibnamefont {Steinhardt}},\
  }\bibfield  {title} {\bibinfo {title} {Optimized structures for photonic
  quasicrystals},\ }\href@noop {} {\bibfield  {journal} {\bibinfo  {journal}
  {Phys. Rev. Lett.}\ }\textbf {\bibinfo {volume} {101}},\ \bibinfo {pages}
  {073902} (\bibinfo {year} {2008})}\BibitemShut {NoStop}%
\bibitem [{\citenamefont {Florescu}\ \emph {et~al.}(2009)\citenamefont
  {Florescu}, \citenamefont {Torquato},\ and\ \citenamefont
  {Steinhardt}}]{florescu2009complete}%
  \BibitemOpen
  \bibfield  {author} {\bibinfo {author} {\bibfnamefont {M.}~\bibnamefont
  {Florescu}}, \bibinfo {author} {\bibfnamefont {S.}~\bibnamefont {Torquato}},\
  and\ \bibinfo {author} {\bibfnamefont {P.~J.}\ \bibnamefont {Steinhardt}},\
  }\bibfield  {title} {\bibinfo {title} {Complete band gaps in two-dimensional
  photonic quasicrystals},\ }\href@noop {} {\bibfield  {journal} {\bibinfo
  {journal} {Phys. Rev. B}\ }\textbf {\bibinfo {volume} {80}},\ \bibinfo
  {pages} {155112} (\bibinfo {year} {2009})}\BibitemShut {NoStop}%
\bibitem [{\citenamefont {Steurer}\ and\ \citenamefont
  {Haibach}(1999)}]{steurer1999periodic}%
  \BibitemOpen
  \bibfield  {author} {\bibinfo {author} {\bibfnamefont {W.}~\bibnamefont
  {Steurer}}\ and\ \bibinfo {author} {\bibfnamefont {T.}~\bibnamefont
  {Haibach}},\ }\bibfield  {title} {\bibinfo {title} {The periodic average
  structure of particular quasicrystals},\ }\href@noop {} {\bibfield  {journal}
  {\bibinfo  {journal} {Acta Crystallogr. A}\ }\textbf {\bibinfo {volume}
  {55}},\ \bibinfo {pages} {48} (\bibinfo {year} {1999})}\BibitemShut {NoStop}%
\bibitem [{\citenamefont {Davies}\ and\ \citenamefont
  {Morini}(2023)}]{davies2023super}%
  \BibitemOpen
  \bibfield  {author} {\bibinfo {author} {\bibfnamefont {B.}~\bibnamefont
  {Davies}}\ and\ \bibinfo {author} {\bibfnamefont {L.}~\bibnamefont
  {Morini}},\ }\bibfield  {title} {\bibinfo {title} {Super band gaps and
  periodic approximants of generalised {Fibonacci} tilings},\ }\href@noop {}
  {\bibfield  {journal} {\bibinfo  {journal} {arXiv preprint arXiv:2302.10063}\
  } (\bibinfo {year} {2023})}\BibitemShut {NoStop}%
\bibitem [{\citenamefont {Mart{\'\i}-Sabat{\'e}}\ and\ \citenamefont
  {Torrent}(2021)}]{marti2021edge}%
  \BibitemOpen
  \bibfield  {author} {\bibinfo {author} {\bibfnamefont {M.}~\bibnamefont
  {Mart{\'\i}-Sabat{\'e}}}\ and\ \bibinfo {author} {\bibfnamefont
  {D.}~\bibnamefont {Torrent}},\ }\bibfield  {title} {\bibinfo {title} {Edge
  modes for flexural waves in quasi-periodic linear arrays of scatterers},\
  }\href@noop {} {\bibfield  {journal} {\bibinfo  {journal} {APL Mater.}\
  }\textbf {\bibinfo {volume} {9}},\ \bibinfo {pages} {081107} (\bibinfo {year}
  {2021})}\BibitemShut {NoStop}%
\bibitem [{\citenamefont {Liu}\ \emph {et~al.}(2022)\citenamefont {Liu},
  \citenamefont {Santos},\ and\ \citenamefont {Prodan}}]{liu2022topological}%
  \BibitemOpen
  \bibfield  {author} {\bibinfo {author} {\bibfnamefont {Y.}~\bibnamefont
  {Liu}}, \bibinfo {author} {\bibfnamefont {L.~F.}\ \bibnamefont {Santos}},\
  and\ \bibinfo {author} {\bibfnamefont {E.}~\bibnamefont {Prodan}},\
  }\bibfield  {title} {\bibinfo {title} {Topological gaps in quasiperiodic spin
  chains: A numerical and {K}-theoretic analysis},\ }\href@noop {} {\bibfield
  {journal} {\bibinfo  {journal} {Phys. Rev. B}\ }\textbf {\bibinfo {volume}
  {105}},\ \bibinfo {pages} {035115} (\bibinfo {year} {2022})}\BibitemShut
  {NoStop}%
\bibitem [{\citenamefont {Ni}\ \emph {et~al.}(2019)\citenamefont {Ni},
  \citenamefont {Chen}, \citenamefont {Weiner}, \citenamefont {Apigo},
  \citenamefont {Prodan}, \citenamefont {Alu}, \citenamefont {Prodan},\ and\
  \citenamefont {Khanikaev}}]{ni2019observation}%
  \BibitemOpen
  \bibfield  {author} {\bibinfo {author} {\bibfnamefont {X.}~\bibnamefont
  {Ni}}, \bibinfo {author} {\bibfnamefont {K.}~\bibnamefont {Chen}}, \bibinfo
  {author} {\bibfnamefont {M.}~\bibnamefont {Weiner}}, \bibinfo {author}
  {\bibfnamefont {D.~J.}\ \bibnamefont {Apigo}}, \bibinfo {author}
  {\bibfnamefont {C.}~\bibnamefont {Prodan}}, \bibinfo {author} {\bibfnamefont
  {A.}~\bibnamefont {Alu}}, \bibinfo {author} {\bibfnamefont {E.}~\bibnamefont
  {Prodan}},\ and\ \bibinfo {author} {\bibfnamefont {A.~B.}\ \bibnamefont
  {Khanikaev}},\ }\bibfield  {title} {\bibinfo {title} {Observation of
  {Hofstadter} butterfly and topological edge states in reconfigurable
  quasi-periodic acoustic crystals},\ }\href@noop {} {\bibfield  {journal}
  {\bibinfo  {journal} {Commun. Phys.}\ }\textbf {\bibinfo {volume} {2}},\
  \bibinfo {pages} {1} (\bibinfo {year} {2019})}\BibitemShut {NoStop}%
\bibitem [{\citenamefont {Kraus}\ \emph {et~al.}(2012)\citenamefont {Kraus},
  \citenamefont {Lahini}, \citenamefont {Ringel}, \citenamefont {Verbin},\ and\
  \citenamefont {Zilberberg}}]{kraus2012topological}%
  \BibitemOpen
  \bibfield  {author} {\bibinfo {author} {\bibfnamefont {Y.~E.}\ \bibnamefont
  {Kraus}}, \bibinfo {author} {\bibfnamefont {Y.}~\bibnamefont {Lahini}},
  \bibinfo {author} {\bibfnamefont {Z.}~\bibnamefont {Ringel}}, \bibinfo
  {author} {\bibfnamefont {M.}~\bibnamefont {Verbin}},\ and\ \bibinfo {author}
  {\bibfnamefont {O.}~\bibnamefont {Zilberberg}},\ }\bibfield  {title}
  {\bibinfo {title} {Topological states and adiabatic pumping in
  quasicrystals},\ }\href@noop {} {\bibfield  {journal} {\bibinfo  {journal}
  {Phys. Rev. Lett.}\ }\textbf {\bibinfo {volume} {109}},\ \bibinfo {pages}
  {106402} (\bibinfo {year} {2012})}\BibitemShut {NoStop}%
\bibitem [{\citenamefont {Prodan}\ and\ \citenamefont
  {Schulz-Baldes}(2016)}]{prodan2016bulk}%
  \BibitemOpen
  \bibfield  {author} {\bibinfo {author} {\bibfnamefont {E.}~\bibnamefont
  {Prodan}}\ and\ \bibinfo {author} {\bibfnamefont {H.}~\bibnamefont
  {Schulz-Baldes}},\ }\href@noop {} {\emph {\bibinfo {title} {Bulk and Boundary
  Invariants for Complex Topological Insulators: From K-Theory to Physics}}}\
  (\bibinfo  {publisher} {Springer},\ \bibinfo {year} {2016})\BibitemShut
  {NoStop}%
\bibitem [{\citenamefont {Davies}\ and\ \citenamefont
  {Craster}(2022)}]{davies2022symmetry}%
  \BibitemOpen
  \bibfield  {author} {\bibinfo {author} {\bibfnamefont {B.}~\bibnamefont
  {Davies}}\ and\ \bibinfo {author} {\bibfnamefont {R.~V.}\ \bibnamefont
  {Craster}},\ }\bibfield  {title} {\bibinfo {title} {Symmetry-induced
  quasicrystalline waveguides},\ }\href@noop {} {\bibfield  {journal} {\bibinfo
   {journal} {Wave Motion}\ }\textbf {\bibinfo {volume} {115}},\ \bibinfo
  {pages} {103068} (\bibinfo {year} {2022})}\BibitemShut {NoStop}%
\bibitem [{\citenamefont {Hofstadter}(1976)}]{hofstadter1976energy}%
  \BibitemOpen
  \bibfield  {author} {\bibinfo {author} {\bibfnamefont {D.~R.}\ \bibnamefont
  {Hofstadter}},\ }\bibfield  {title} {\bibinfo {title} {Energy levels and wave
  functions of {Bloch} electrons in rational and irrational magnetic fields},\
  }\href@noop {} {\bibfield  {journal} {\bibinfo  {journal} {Phys. Rev. B}\
  }\textbf {\bibinfo {volume} {14}},\ \bibinfo {pages} {2239} (\bibinfo {year}
  {1976})}\BibitemShut {NoStop}%
\bibitem [{\citenamefont {Pal}\ \emph {et~al.}(2019)\citenamefont {Pal},
  \citenamefont {Rosa},\ and\ \citenamefont {Ruzzene}}]{pal2019topological}%
  \BibitemOpen
  \bibfield  {author} {\bibinfo {author} {\bibfnamefont {R.~K.}\ \bibnamefont
  {Pal}}, \bibinfo {author} {\bibfnamefont {M.~I.~N.}\ \bibnamefont {Rosa}},\
  and\ \bibinfo {author} {\bibfnamefont {M.}~\bibnamefont {Ruzzene}},\
  }\bibfield  {title} {\bibinfo {title} {Topological bands and localized
  vibration modes in quasiperiodic beams},\ }\href@noop {} {\bibfield
  {journal} {\bibinfo  {journal} {New J. Phys.}\ }\textbf {\bibinfo {volume}
  {21}},\ \bibinfo {pages} {093017} (\bibinfo {year} {2019})}\BibitemShut
  {NoStop}%
\bibitem [{\citenamefont {Apigo}\ \emph {et~al.}(2018)\citenamefont {Apigo},
  \citenamefont {Qian}, \citenamefont {Prodan},\ and\ \citenamefont
  {Prodan}}]{apigo2018topological}%
  \BibitemOpen
  \bibfield  {author} {\bibinfo {author} {\bibfnamefont {D.~J.}\ \bibnamefont
  {Apigo}}, \bibinfo {author} {\bibfnamefont {K.}~\bibnamefont {Qian}},
  \bibinfo {author} {\bibfnamefont {C.}~\bibnamefont {Prodan}},\ and\ \bibinfo
  {author} {\bibfnamefont {E.}~\bibnamefont {Prodan}},\ }\bibfield  {title}
  {\bibinfo {title} {Topological edge modes by smart patterning},\ }\href@noop
  {} {\bibfield  {journal} {\bibinfo  {journal} {Phys. Rev. Mater.}\ }\textbf
  {\bibinfo {volume} {2}},\ \bibinfo {pages} {124203} (\bibinfo {year}
  {2018})}\BibitemShut {NoStop}%
\bibitem [{\citenamefont {Wilks}\ \emph {et~al.}(2022)\citenamefont {Wilks},
  \citenamefont {Montiel},\ and\ \citenamefont {Wakes}}]{wilks2022rainbow}%
  \BibitemOpen
  \bibfield  {author} {\bibinfo {author} {\bibfnamefont {B.}~\bibnamefont
  {Wilks}}, \bibinfo {author} {\bibfnamefont {F.}~\bibnamefont {Montiel}},\
  and\ \bibinfo {author} {\bibfnamefont {S.}~\bibnamefont {Wakes}},\ }\bibfield
   {title} {\bibinfo {title} {Rainbow reflection and broadband energy
  absorption of water waves by graded arrays of vertical barriers},\ }\href
  {https://doi.org/10.1017/jfm.2022.302} {\bibfield  {journal} {\bibinfo
  {journal} {J. Fluid Mech.}\ }\textbf {\bibinfo {volume} {941}},\ \bibinfo
  {pages} {A26} (\bibinfo {year} {2022})}\BibitemShut {NoStop}%
\bibitem [{\citenamefont {Rosafalco}\ \emph {et~al.}(2023)\citenamefont
  {Rosafalco}, \citenamefont {De~Ponti}, \citenamefont {Iorio}, \citenamefont
  {Ardito},\ and\ \citenamefont {Corigliano}}]{rosafalco2022optimised}%
  \BibitemOpen
  \bibfield  {author} {\bibinfo {author} {\bibfnamefont {L.}~\bibnamefont
  {Rosafalco}}, \bibinfo {author} {\bibfnamefont {J.~M.}\ \bibnamefont
  {De~Ponti}}, \bibinfo {author} {\bibfnamefont {L.}~\bibnamefont {Iorio}},
  \bibinfo {author} {\bibfnamefont {R.}~\bibnamefont {Ardito}},\ and\ \bibinfo
  {author} {\bibfnamefont {A.}~\bibnamefont {Corigliano}},\ }\bibfield  {title}
  {\bibinfo {title} {Optimised graded metamaterials for mechanical energy
  confinement and amplification via reinforcement learning},\ }\href@noop {}
  {\bibfield  {journal} {\bibinfo  {journal} {Eur. J. Mech. A Solids}\ }\textbf
  {\bibinfo {volume} {99}},\ \bibinfo {pages} {104947} (\bibinfo {year}
  {2023})}\BibitemShut {NoStop}%
\bibitem [{\citenamefont {Cebrecos}\ \emph {et~al.}(2014)\citenamefont
  {Cebrecos}, \citenamefont {Picó}, \citenamefont {Sánchez-Morcillo},
  \citenamefont {Staliunas}, \citenamefont {Romero-García},\ and\
  \citenamefont {Garcia-Raffi}}]{rg_exponential}%
  \BibitemOpen
  \bibfield  {author} {\bibinfo {author} {\bibfnamefont {A.}~\bibnamefont
  {Cebrecos}}, \bibinfo {author} {\bibfnamefont {R.}~\bibnamefont {Picó}},
  \bibinfo {author} {\bibfnamefont {V.~J.}\ \bibnamefont {Sánchez-Morcillo}},
  \bibinfo {author} {\bibfnamefont {K.}~\bibnamefont {Staliunas}}, \bibinfo
  {author} {\bibfnamefont {V.}~\bibnamefont {Romero-García}},\ and\ \bibinfo
  {author} {\bibfnamefont {L.~M.}\ \bibnamefont {Garcia-Raffi}},\ }\bibfield
  {title} {\bibinfo {title} {Enhancement of sound by soft reflections in
  exponentially chirped crystals},\ }\href {https://doi.org/10.1063/1.4902508}
  {\bibfield  {journal} {\bibinfo  {journal} {AIP Adv.}\ }\textbf {\bibinfo
  {volume} {4}},\ \bibinfo {pages} {124402} (\bibinfo {year} {2014})},\ \Eprint
  {https://arxiv.org/abs/https://doi.org/10.1063/1.4902508}
  {https://doi.org/10.1063/1.4902508} \BibitemShut {NoStop}%
\bibitem [{\citenamefont {Davies}\ \emph {et~al.}(2023)\citenamefont {Davies},
  \citenamefont {Fehertoi-Nagy},\ and\ \citenamefont
  {Putley}}]{davies2023difficulty}%
  \BibitemOpen
  \bibfield  {author} {\bibinfo {author} {\bibfnamefont {B.}~\bibnamefont
  {Davies}}, \bibinfo {author} {\bibfnamefont {L.}~\bibnamefont
  {Fehertoi-Nagy}},\ and\ \bibinfo {author} {\bibfnamefont {H.}~\bibnamefont
  {Putley}},\ }\bibfield  {title} {\bibinfo {title} {On the problem of
  comparing graded metamaterials},\ }\href@noop {} {\bibfield  {journal}
  {\bibinfo  {journal} {arXiv preprint arXiv:2305.00904}\ } (\bibinfo {year}
  {2023})}\BibitemShut {NoStop}%
\bibitem [{\citenamefont {Chaplain}\ \emph
  {et~al.}(2020{\natexlab{a}})\citenamefont {Chaplain}, \citenamefont
  {De~Ponti}, \citenamefont {Aguzzi}, \citenamefont {Colombi},\ and\
  \citenamefont {Craster}}]{chaplain2020topological}%
  \BibitemOpen
  \bibfield  {author} {\bibinfo {author} {\bibfnamefont {G.~J.}\ \bibnamefont
  {Chaplain}}, \bibinfo {author} {\bibfnamefont {J.~M.}\ \bibnamefont
  {De~Ponti}}, \bibinfo {author} {\bibfnamefont {G.}~\bibnamefont {Aguzzi}},
  \bibinfo {author} {\bibfnamefont {A.}~\bibnamefont {Colombi}},\ and\ \bibinfo
  {author} {\bibfnamefont {R.~V.}\ \bibnamefont {Craster}},\ }\bibfield
  {title} {\bibinfo {title} {Topological rainbow trapping for elastic energy
  harvesting in graded {Su-Schrieffer-Heeger} systems},\ }\href@noop {}
  {\bibfield  {journal} {\bibinfo  {journal} {Phys. Rev. Appl.}\ }\textbf
  {\bibinfo {volume} {14}},\ \bibinfo {pages} {054035} (\bibinfo {year}
  {2020}{\natexlab{a}})}\BibitemShut {NoStop}%
\bibitem [{\citenamefont {Chaplain}\ \emph
  {et~al.}(2020{\natexlab{b}})\citenamefont {Chaplain}, \citenamefont {Pajer},
  \citenamefont {De~Ponti},\ and\ \citenamefont
  {Craster}}]{chaplain2020delineating}%
  \BibitemOpen
  \bibfield  {author} {\bibinfo {author} {\bibfnamefont {G.~J.}\ \bibnamefont
  {Chaplain}}, \bibinfo {author} {\bibfnamefont {D.}~\bibnamefont {Pajer}},
  \bibinfo {author} {\bibfnamefont {J.~M.}\ \bibnamefont {De~Ponti}},\ and\
  \bibinfo {author} {\bibfnamefont {R.~V.}\ \bibnamefont {Craster}},\
  }\bibfield  {title} {\bibinfo {title} {Delineating rainbow reflection and
  trapping with applications for energy harvesting},\ }\href@noop {} {\bibfield
   {journal} {\bibinfo  {journal} {New J. Phys.}\ }\textbf {\bibinfo {volume}
  {22}},\ \bibinfo {pages} {063024} (\bibinfo {year}
  {2020}{\natexlab{b}})}\BibitemShut {NoStop}%
\bibitem [{\citenamefont {Janot}(1994)}]{janot1994quasicrystals}%
  \BibitemOpen
  \bibfield  {author} {\bibinfo {author} {\bibfnamefont {C.}~\bibnamefont
  {Janot}},\ }\href@noop {} {\emph {\bibinfo {title} {Quasicrystals}}},\
  \bibinfo {edition} {2nd}\ ed.\ (\bibinfo  {publisher} {Clarendon},\ \bibinfo
  {address} {Oxford},\ \bibinfo {year} {1994})\BibitemShut {NoStop}%
\bibitem [{\citenamefont {Rodriguez}\ \emph {et~al.}(2008)\citenamefont
  {Rodriguez}, \citenamefont {McCauley}, \citenamefont {Avniel},\ and\
  \citenamefont {Johnson}}]{rodriguez2008computation}%
  \BibitemOpen
  \bibfield  {author} {\bibinfo {author} {\bibfnamefont {A.~W.}\ \bibnamefont
  {Rodriguez}}, \bibinfo {author} {\bibfnamefont {A.~P.}\ \bibnamefont
  {McCauley}}, \bibinfo {author} {\bibfnamefont {Y.}~\bibnamefont {Avniel}},\
  and\ \bibinfo {author} {\bibfnamefont {S.~G.}\ \bibnamefont {Johnson}},\
  }\bibfield  {title} {\bibinfo {title} {Computation and visualization of
  photonic quasicrystal spectra via {Bloch’s} theorem},\ }\href@noop {}
  {\bibfield  {journal} {\bibinfo  {journal} {Phys. Rev. B}\ }\textbf {\bibinfo
  {volume} {77}},\ \bibinfo {pages} {104201} (\bibinfo {year}
  {2008})}\BibitemShut {NoStop}%
\bibitem [{\citenamefont {Amenoagbadji}\ \emph {et~al.}(2023)\citenamefont
  {Amenoagbadji}, \citenamefont {Fliss},\ and\ \citenamefont
  {Joly}}]{amenoagbadji2023wave}%
  \BibitemOpen
  \bibfield  {author} {\bibinfo {author} {\bibfnamefont {P.}~\bibnamefont
  {Amenoagbadji}}, \bibinfo {author} {\bibfnamefont {S.}~\bibnamefont
  {Fliss}},\ and\ \bibinfo {author} {\bibfnamefont {P.}~\bibnamefont {Joly}},\
  }\bibfield  {title} {\bibinfo {title} {Wave propagation in one-dimensional
  quasiperiodic media},\ }\href@noop {} {\bibfield  {journal} {\bibinfo
  {journal} {arXiv preprint arXiv:2301.01159}\ } (\bibinfo {year}
  {2023})}\BibitemShut {NoStop}%
\bibitem [{\citenamefont {{COMSOL Multiphysics{\textregistered} v.
  6.0}}()}]{Comsol}%
  \BibitemOpen
  \bibfield  {author} {\bibinfo {author} {\bibnamefont {{COMSOL
  Multiphysics{\textregistered} v. 6.0}}},\ }\href@noop {} {\emph {\bibinfo
  {title} {\href{www.comsol.com/}{www.comsol.com/}}}}\ (\bibinfo  {publisher}
  {COMSOL AB},\ \bibinfo {address} {Stockholm, Sweden})\BibitemShut {NoStop}%
\end{thebibliography}%

\clearpage
\onecolumngrid

\begin{center}
    \textbf{\large Supplementary Materials}
\end{center}

\section{Theoretical ODE Model}

Here, we give details of the simple one-dimensional ordinary-differential model that was used to make theoretical predictions about the qualitative behaviour of the system.

We assume time-harmonic excitation with frequency $\omega$, in which case the time-harmonic solution to the scalar wave equation $u(x)$ satisfies the ordinary differential equation
\begin{equation} \label{eq:helmholtz}
    \frac{\mathrm{d}^2u}{\mathrm{d}x^2} + \omega^2 \frac{\rho^2}{Z^2} u=0,
\end{equation}
where $\rho$ is the density and $Z$ is the characteristic acoustic impedance. We suppose that $\rho=1$~kg~m$^{-3}$ does not vary, since the waveguide contains air throughout. The acoustic impedance $Z=Z(x)$ is assumed to vary, such that it increases each time the wave passes one of the scatterers. The positions of the scatterers within the waveguide are given by the curved cut-and-project algorithm outlined in the main body of the article. This algorithm gives us as set of points $\P(\lambda)\subset\R$ at which we should place the barriers. Then, we take $Z$ to have the form
\begin{equation}
    Z(x)=
    \begin{cases}
        Z_s & \text{if } \min\limits_{z\in\P(\Lambda)}|x-z|<l, \\
        Z_0 &\text{otherwise},
    \end{cases}
\end{equation}
where $Z_0$ and $Z_s$ are constants. In our experimental setup, the scatterers are 4~mm-diameter rods placed in the centre of the square waveguide with 2~cm cross section. For our theoretical results we take $l=4$~mm, and use the impedance values $Z_0=5.3\times10^4$~Pa~s~m$^{-1}$ and $Z_s=7.2\times10^4$~Pa~s~m$^{-1}$. This gives a contrast of $Z_s/Z_0=1.35$. These parameter values were chosen to align the 8~kHz band gap in the transmission spectra with the FE simulations (Fig.~3 of the main article). In the setting of the one-dimensional model \eqref{eq:helmholtz}, little attention should be paid to the choice of parameter values, it is only the qualitative behaviour that is of interest for informing subsequent FE simulations and experiments.

We will study the solution to \eqref{eq:helmholtz} using transfer matrices. For an $[a,b)\subset\R$ we define $T_{x,y}^\omega\in\mathbb{R}^{2\times2}$ to be the matrix satisfying
\begin{equation}
\begin{pmatrix} u(b) \\ u'(b) \end{pmatrix} = T_{a,b}^\omega \begin{pmatrix} u(a) \\ u'(a) \end{pmatrix}.
\end{equation}
That is, given the solution and its derivative at $a$, multiplication by $T_{a,b}^\omega$ gives the solution and its derivative at $b$. Since $Z(x)$ is piecewise constant, it is straightforward to calculate that for any interval $[a,b)$ on which $Z$ is constant
\begin{equation}
T_{a,b}^\omega=\begin{pmatrix}
\cos\big(\frac{\omega\rho(b-a)}{Z}\big) & \frac{1}{\omega}\frac{Z}{\rho}\sin\big(\frac{\omega\rho(b-a)}{Z}\big) \\
-\omega\frac{\rho}{Z} \sin\big(\frac{\omega\rho(b-a)}{Z}\big) & \cos\big(\frac{\omega\rho(b-a)}{Z}\big)
\end{pmatrix}.
\end{equation}

\begin{figure}
    \centering
    \includegraphics[width=0.95\linewidth,clip,trim=1.5cm 0 2cm 0]{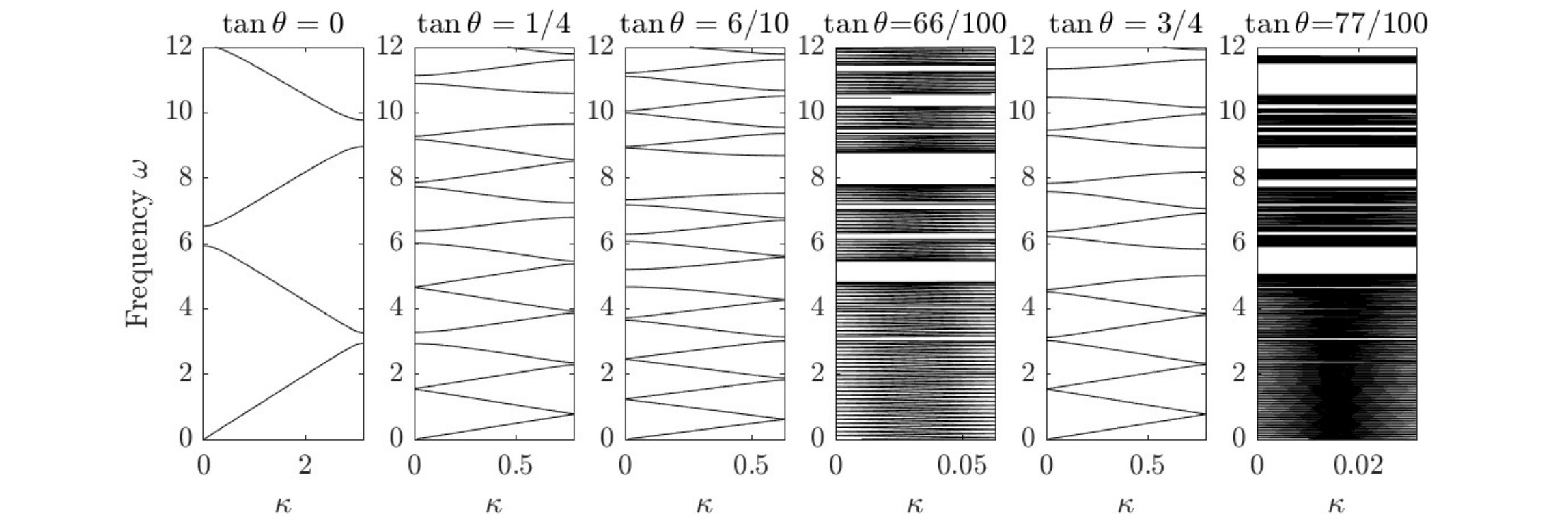}
    \caption{If the projection angle $\theta$ is such that $\tan\theta$ is rational, then the projected quasicrystal is periodic so its spectrum can be computed using Floquet-Bloch.}
    \label{fig:bands}
\end{figure}

The spectra of quasicrystals obtained by projecting onto a straight line $\Gamma$ which is at an angle $\theta$ can be computed straightforwardly if $\theta$ is such that $\tan\theta$ is rational, since the projected crystals are periodic in this case. In particular, if $\tan\theta=m/n$, then the projected crystal has a unit cell of length
\begin{equation} \label{eq:length}
L=\frac{n}{\gcd(n,m)},
\end{equation}
where $\gcd(n,m)$ is the greatest common divisor of $n$ and $m$. The Bloch ansatz is that $u(x+L)=e^{\I\kappa L}u(x)$ for some real valued $\kappa$. This leads to the equation
\begin{equation}
\det\left(T_{0,L}^\omega-e^{\I\kappa L}I\right)=0.
\end{equation}
Using the fact that $\det(T_{0,L}^\omega)=1$, we can derive the dispersion relation
\begin{equation}
2\cos(\kappa L) = \tr(T_{0,L}^\omega).
\end{equation}
This has real solutions for $\kappa$ if and only if $|\tr(T_{0,L}^\omega)|\leq2$, meaning that $\omega$ is in a band gap when $|\tr(T_{0,L}^\omega)|>2$. For example, the dispersion curves for some selected rational values of $\tan\theta$ are shown in Fig.~\ref{fig:bands}. Notice that in each case we plot the Bloch parameter in the range $0\leq\kappa\leq \pi/L$ on the horizontal axes, where $L$ varies for different values of $\theta$, as specified in the formula \eqref{eq:length}. The ``butterfly'' shown in Fig.~2 of the main article is generated by computing the Bloch dispersion curves for 200 evenly spaced rational values of $\tan\theta$ between 0 and 2.

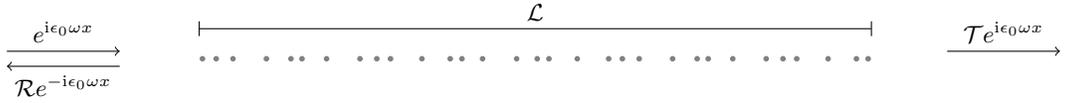
\begin{figure}
\centering
\begin{tikzpicture}
\foreach \x in {0,1,2,...,30}{
\draw[gray,fill=gray] ({0.3*\x+0.1*cos(100*\x)},0) circle (0.03);
}
\draw[|-|] (0.05,0.4) -- (9,0.4) node[above,midway]{$\L$};
\draw[->] (-2.5,0.1) -- (-1,0.1) node[above,midway]{$e^{\I\epsilon_0\omega x}$};
\draw[<-] (-2.5,-0.1) -- (-1,-0.1) node[below,midway]{$\rr e^{-\I\epsilon_0\omega x}$};
\draw[->] (10,0.1) -- (11.5,0.1) node[above,midway]{$\tt e^{\I\epsilon_0\omega x}$};
\end{tikzpicture}
\caption{For a finite-sized array of length $\L$, we calculate the reflectance $|\rr|^2$ and the transmittance $|\tt|^2$ for different frequencies $\omega$.} \label{fig:tandr}
\end{figure}

Given a finite-sized piece of material of length $\L$, the reflectance and transmittance can be calculated using the transfer matrix $T_{0,\L}(\omega)$. We suppose that a wave $\exp(\I\rho\omega x/Z_0)$ is incident on one side of the array. Then, there will be a reflected wave $\rr\exp(-\I\rho\omega x/Z_0)$ and a transmitted wave $\tt\exp(\I\rho\omega x/Z_0)$, where $\rr,\tt\in\C$ (this is depicted in Fig.~\ref{fig:tandr}). Hence, if the array occupies the region $[0,\L]$, we have that
\begin{equation}
u(x)=\begin{cases}
e^{\I\rho\omega x/Z_0} + \rr e^{-\I\rho\omega x/Z_0} & \text{for } x\leq0, \\
\tt e^{\I\rho\omega x/Z_0} & \text{for } x\geq \L.
\end{cases}
\end{equation}
Now, we have also that 
\begin{equation}
\begin{pmatrix} u(\L) \\ u'(\L) \end{pmatrix} = T_{0,\L}(\omega) \begin{pmatrix} u(0) \\ u'(0) \end{pmatrix},
\end{equation}
from which we are able to solve for the constants $\rr$ and $\tt$. We find that 
\begin{gather}
\rr = \frac{(T_{0,\L})_{21}+\Omega_0^2(T_{0,\L})_{12}-\I\Omega_0(T_{0,\L})_{11}+\I\Omega_0(T_{0,\L})_{22}}{-(T_{0,\L})_{21}+\Omega_0^2(T_{0,\L})_{12}+\I\Omega_0(T_{0,\L})_{11}+\I\Omega_0(T_{0,\L})_{22}}, \\
\tt = \frac{2\I\Omega_0 \exp(-\I\Omega_0\L)}{-(T_{0,\L})_{21}+\Omega_0^2(T_{0,\L})_{12}+\I\Omega_0(T_{0,\L})_{11}+\I\Omega_0(T_{0,\L})_{22}},
\end{gather}
where the notation $\Omega_0=\rho\omega/Z_0$ has been introduced for brevity. It is easy to check that $|\rr|^2+|\tt|^2=1$, showing that energy is conserved. The quantity $t=|\tt|^2$ is the transmittance, plotted in Fig.~3 of the main article.

\section{Experimental Setup \& Numerical Modelling}

Here, we further detail the experimental configuration and positional measurements that confirm the existence of the fractal rainbow in the main text. 

\begin{figure}
    \centering
    \includegraphics[width = 0.9\textwidth]{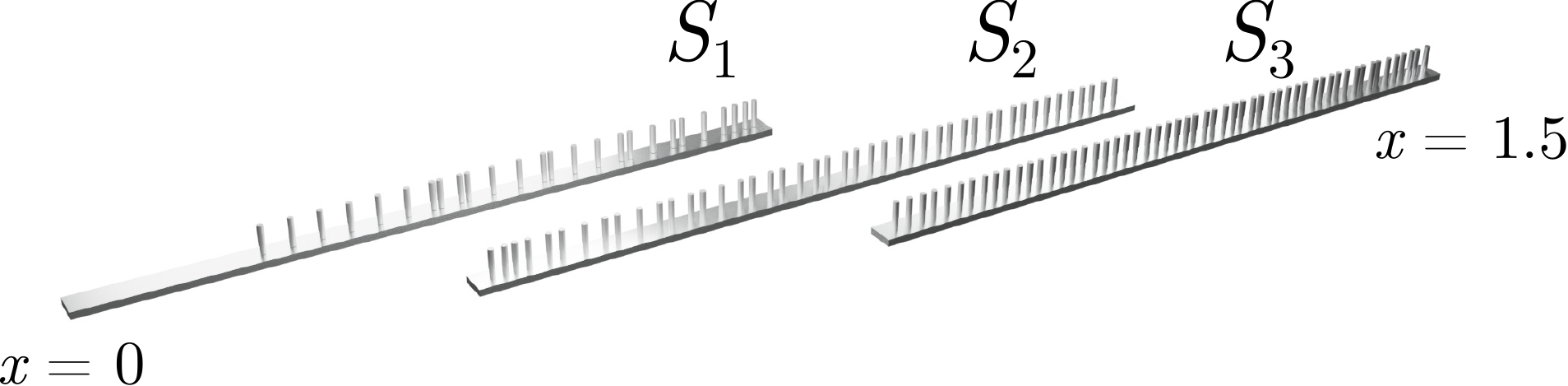}
    \caption{Schematic of sections $S_i$ forming the graded quasi-periodic array. The centre of the scatterer positions in each section are shown in Table 1.}
    \label{fig:SectionSchem}
\end{figure}

Figure~\ref{fig:SectionSchem} shows a schematic of the bottom plate (and rods) that, for convenience, is split into three sections $S_i$, $i$ = 1,2,3. Circular holes are milled through the bottom plate at the positions ($x$-coordinates) detailed in Table~\ref{tab:table3}. The rods are then inserted into the holes such that the centre of the rods' circular cross section lies along $y = 0$. The end of $S_1$, $x = 0$, is the entrance of the waveguide in Fig. 1 in the main text. 

\begin{figure}
    \centering
    \includegraphics[width = 0.95\textwidth]{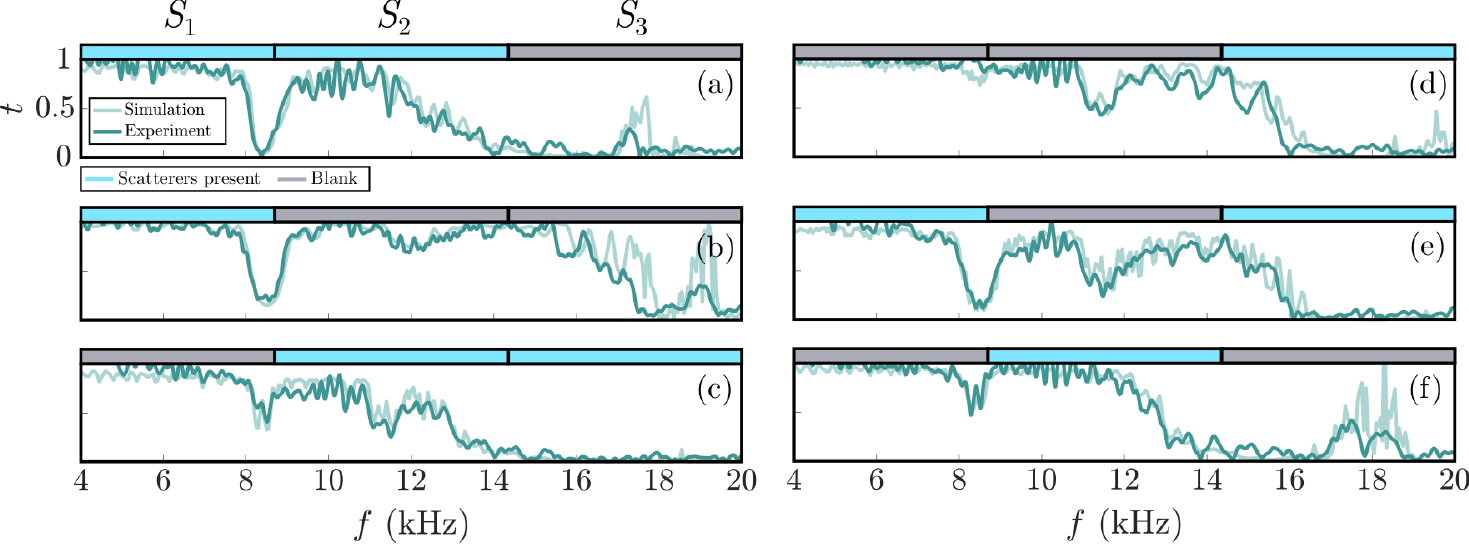}
    \caption{Comparisons between experimental results and (2D, lossy) FE simulations for different combinations of sections $S_i$ where scatterers are present, shown by the schematic above each panel with blue representing the presence of the rods and purple showing their absence. We write this in short hand with $S$ for scatterers present and $B$ for a blank section. The combinations are thus (a) $SSB$, (b) $SBB$, (c) $BSS$, (d) $BBS$, (e) $SBS$, (f) $BSB$. }
    \label{fig:Sections}
\end{figure}

In additional to convenience for assembly, the `modular' nature of the sections $S_i$ allows a further investigation into how the configuration of the graded quasicrystal affects the local band structure. In Fig.~\ref{fig:Sections} we show six additional comparisons between experiment and (lossy) FE simulations for the permutations of each of the sections with and without the scatterers. We adopt the notation for each section to be $S$ if the rods are present, and $B$ (blank) for when they are absent. As such we denote the configurations $SSB$, $SBB$, $BSS$, $BBS$, $SBS$, $BSB$. The final two configurations $SSS$ and $BBB$ are shown in the main text and used as the reference, respectively. We show these permutations above the transmission plots (akin to Fig 3.) in Fig.~\ref{fig:Sections} schematically, as blue/purple blocks when the scatterers are present/absent. 

The effect of each of the sections is clear: each section inherits their own local gaps due to the underlying periodic structure from which the quasi-crystal is projected, with the gaps around 8~kHz, 14~kHz and 16~kHz resulting from the distribution of the average separation distances in each section. Convincing agreement is seen between all simulations and experiments.

\begin{figure}
    \centering
    \includegraphics[width = 0.8\textwidth]{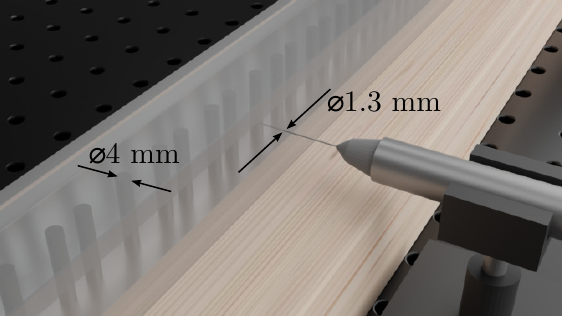}
    \caption{Schematic of needle microphone used for positional measurements (Fig. 4 in the main text). Small holes of diameter 1.3 mm are milled into one side of the waveguide hood, at distances of $x$  = 0.41, 0.58, 0.76 0.94 m. }
    \label{fig:PosSchem}
\end{figure}

In Fig.~\ref{fig:PosSchem} we show a detailed schematic of the setup used in the positional measurements (Fig 4. in the main text). Four small holes (diameter 1.3 mm) are drilled in the side of the hood (made transparent in Fig.~\ref{fig:PosSchem}) with a needle microphone (diameter 1.25 mm) inserted  halfway between the wall of the waveguide and the centre of the scatterers. The experiment is then performed as in the normal transmission setup with and without the rods ($SSS$ and $BBB$ configurations), and the relative ratio of Fourier amplitudes (experimentally) and pressures (numerically) being compared in Fig. 4 in the main text, giving $t$. 

Here, we additionally detail the Finite Element (FE) modelling.

\begin{figure}
    \centering
    \includegraphics[width = 0.75\textwidth]{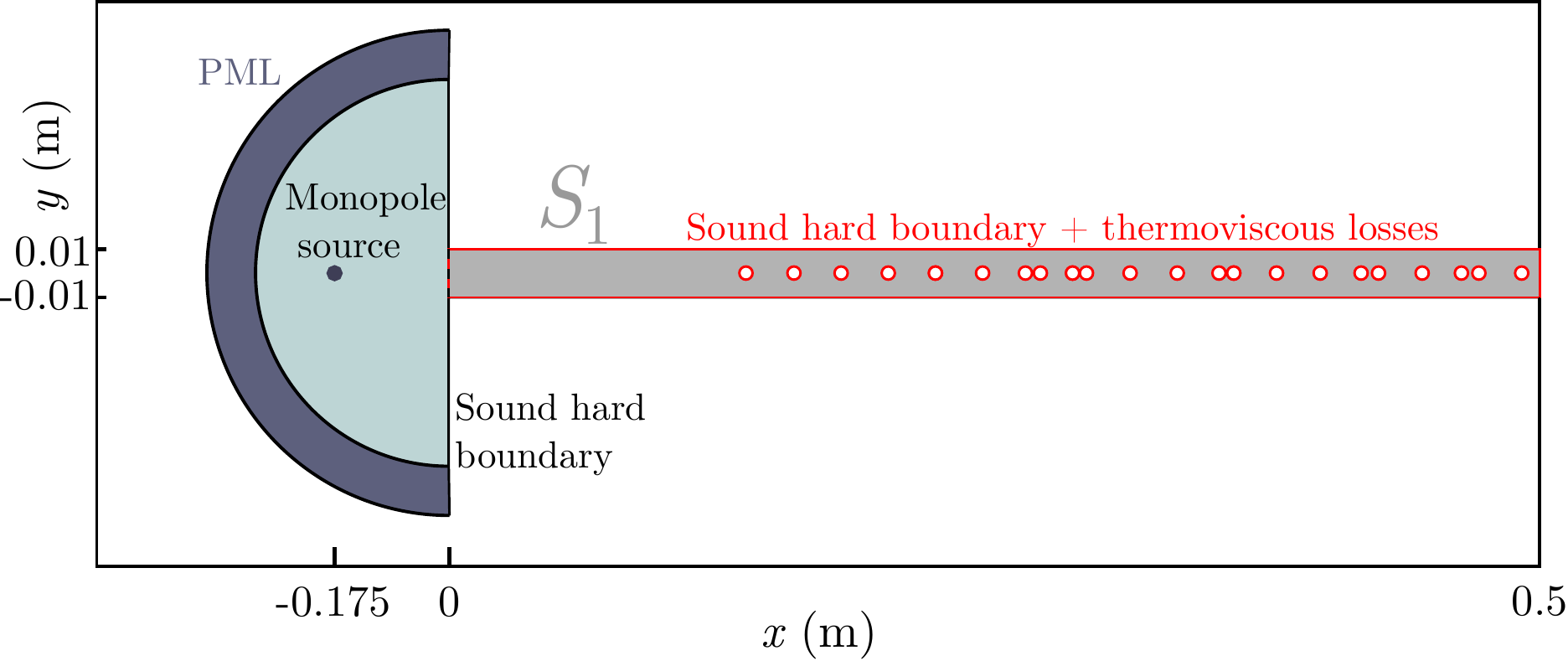}
    \caption{Schematic of portion of finite element model. Perfectly matched layers are shown, along with the source position. Section $S_1$ is shown with scatterers present, modelled as voids with thermoviscous boundaries; a reference simulation is also performed without the scatterers present to calculate $t$. The multiphysics interface is shown by the dashed red boundary at the waveguide entrance ($x = 0$ m).}
    \label{fig:FemSchem}
\end{figure}

Frequency domain simulations are performed, in 2D, using Comsol Multiphysics incorporating thermoviscous acoustics within the waveguide \cite{Comsol}. A schematic of a section of the computational domain is shown in Fig.~\ref{fig:FemSchem}, showing the first section $S_1$ with the scatterers modelled as voids with thermovicous boundary layers. A numerical version of the experiment is conducted comparing the pressure amplitude with and without the scatterers present, at the microphone position at $x = 1.5$ m. The loudspeaker source is approximated by a monopole point source at a distance $-175$ mm from the entry to the waveguide (at $x = 0$). Perfectly matched layers surround the source and are also employed at the termination of the waveguide.

\begin{table}
\caption{\label{tab:table3} Scatterer line positions.}
\begin{tabularx}{0.255\textwidth}{ccc}
& $x$-coordinate (m)  & 
\\ \hline
  $S_1$ & $S_2$ & $S_3$ \\
 \hline
0.1222 &  0.4978 &   0.9991    \\
0.1417 &  0.5066       &   1.0097      \\
0.1612 &  0.5137      &   1.0204      \\
0.1807 &  0.5226       &   1.0292      \\
0.2002 &  0.5368       &   1.0399     \\
0.2197 &  0.5456       &   1.0505     \\
0.2374 &  0.5616       &   1.0593     \\
0.2427 &  0.5757       &   1.0700      \\
0.2569 &  0.5846       &   1.0788      \\
0.2622 &  0.6005       &   1.0877      \\
0.2799 &  0.6147       &   1.0983      \\
0.2994 &  0.6236       &   1.1072      \\
0.3171 &  0.6377       &   1.1160      \\
0.3224 &  0.6466       &   1.1267      \\
0.3401 &  0.6608       &   1.1355      \\
0.3578 &  0.6749       &   1.1444      \\
0.3756 &  0.6838       &   1.1532      \\
0.3826 &  0.6980       &   1.1639      \\
0.4004 &  0.7068       &   1.1727      \\
0.4163 &  0.7210       &   1.1816      \\
0.4234 &  0.7334       &   1.1904      \\
0.4411 &  0.7422       &   1.1993      \\
0.4570 &  0.7564       &   1.2099      \\
0.4659 &  0.7653       &   1.2188      \\
0.4748 &  0.7777       &   1.2276      \\
0.4818 &  0.7883       &   1.2365      \\
       &  0.8007       &   1.2453      \\
       &  0.8113       &   1.2524      \\
       &  0.8220       &   1.2631      \\
       &  0.8344       &   1.2702      \\
       &  0.8432       &   1.2808      \\
       &  0.8556       &   1.2896     \\
       &  0.8645       &   1.2967     \\
       &  0.8769       &   1.3074      \\
       &  0.8857       &   1.3144      \\
       &  0.8981       &   1.3251      \\
       &  0.9070       &   1.3322      \\
       &  0.9176       &   1.3410      \\
       &  0.9283       &   1.3499      \\
       &  0.9389       &   1.3570      \\
       &  0.9495       &   1.3676      \\
       &  0.9601       &   1.3747     \\
       &  0.9690      &   1.3835      \\
       &  0.9796       &   1.3906      \\
       &  0.9903       &   1.3995\\
       &         &   1.4101 \\
       &         &   1.4154 \\
       &         &   1.4260\\
       &         &   1.4314\\
       &         &   1.4420\\
       &         &   1.4508\\
       &         &   1.4579\\
      &          &   1.4668\\
      &          &   1.4721\\
      &          &   1.4827\\
      &          &   1.4880\\ 
\end{tabularx}
\end{table}

\end{document}